\documentclass[twocolumn,showpacs,preprintnumbers,amsmath,amssymb,prd,letterpaper,floatfix,nofootinbib,superscriptaddress]{revtex4-1}  
\usepackage{graphicx}
\usepackage{epsfig}
\usepackage{color}
\usepackage{longtable}
\usepackage{hyperref}
\usepackage{latexsym}
\usepackage{amsmath}
\usepackage{amssymb}
\usepackage{dsfont}
\usepackage{verbatim}
\usepackage{bm}

\newcommand{\FC}{\;,}
\newcommand{\FD}{\;.}
\newcommand{\I}{\mathrm{i}}  
\newcommand{\E}{\mathrm{e}}  

\begin{document}


\date{\today}
\title{$K\pi$ scattering and the $K^*$ decay width from lattice QCD}



\vspace{0.2cm}

\author{Sasa Prelovsek}
\email{sasa.prelovsek@ijs.si}
\affiliation{Department of Physics, University of Ljubljana, Slovenia}
\affiliation{Jozef Stefan Institute, Ljubljana, Slovenia}

\author{Luka Leskovec}
\email{luka.leskovec@ijs.si}
\affiliation{Jozef Stefan Institute, Ljubljana, Slovenia}

\author{C. B. Lang}
\email{christian.lang@uni-graz.at}
\affiliation{Institut f\"ur Physik,  University of Graz, A--8010 Graz, Austria}

\author{Daniel Mohler}
\email{dmohler@fnal.gov}
\affiliation{Fermi National Accelerator Laboratory, Batavia, Illinois, USA\\
}

\begin{abstract}
$K^*$ mesons and in particular the $K^*(892)$ were frequently addressed in lattice simulations, but always while ignoring that the $K^*(892)$  decays strongly. We present an exploratory extraction of the  masses and widths for the $K^*$ resonances by simulating $K\pi$ scattering in $p$-wave with $I=1/2$ on the lattice. The $K\pi$ system with total momenta $P=\tfrac{2\pi}{L}e_z~,\ \tfrac{2\pi}{L}(e_x+e_y)$ and $0$, that allows the extraction of  phase shifts at several values of $K\pi$ relative momenta, is studied.
A Breit-Wigner fit of the phase renders a $K^*(892)$ resonance mass $m^{lat}=891\pm 14~$MeV and the $K^*(892)\to K\pi$ coupling $g^{lat}=5.7\pm 1.6$ compared to the experimental values $m^{exp}\approx 892$ MeV and  $g^{exp}=5.72\pm 0.06$, where $g$ parametrizes the $K^*\to K\pi$ width. 
When extracting the phase shift around the $K^*(1410)$ and $K_2^*(1430)$ resonances we take into account the mixing of $p$-wave with $d$-wave  and assume that the scattering is elastic in our simulation. 
This gives us an estimate of  the $K^*(1410)$ resonance mass $m^{lat}=1.33\pm 0.02 ~$GeV compared to $m^{exp}=1.414\pm 0.0015~$GeV assuming the experimental $K^*(1410)\to K\pi$ coupling.  We contrast the resonant $I=1/2$ channel with the repulsive non-resonant $I=3/2$ channel, where the phase is found to be negative and small, in agreement with experiment.  
\end{abstract}

\maketitle

\section{Introduction}\label{sec_introduction}

The $K^*(892)$ meson was addressed in numerous lattice simulations ranging from spectroscopy to weak matrix elements, where $K^*$ appears in the final state. Yet all previous simulations assumed the so-called narrow width approximation, where $K^*$ is assumed to be stable against strong decay and 
the lattice energy level $E$ is simply identified with the $K^*$ energy $\sqrt{p^{\;2}+m_{K*}^2}$. In nature, however, the $K^*(892)$ strongly decays exclusively to $K\pi$ with a rather narrow decay width $\Gamma\simeq 50~$MeV due to the small phase space related to the near-by threshold.  So the asymptotic state in a lattice simulation is not $K^*(892)$ but rather a scattering state $K(p_K)\pi(p_\pi)$.  

The lowest scattering level with total momentum $P=0$ allowed by angular conservation $K(1)\pi(-1)$\footnote{The momenta in parenthesis will be sometimes given in units of $2\pi/L$.} has relatively high energy $E_{sc}\simeq \sqrt{m_\pi^2+(\tfrac{2\pi}{L})^2}+\sqrt{m_K^2+(\tfrac{2\pi}{L})^2}$ and has a rather insignificant effect for typical $m_\pi>m_\pi^{phy}$ and  $L<3~$fm used in most of previous simulations. 
As state-of-the-art simulations aim at physical $m_\pi$ and consequently large $L$, the $K^*\to K\pi$ strong decay is even more influential and is limiting the precision for extraction of phenomenologically important quantities at present  (for example $B\to K^*$ and $D\to K^*$ form factors \cite{Liu:2011raa,Becirevic:2006nm}).

The aim of our paper is to make the first exploratory lattice investigation to address the unstable nature of the ground state resonance $K^*(892)$  as well as the excited resonance $K^*(1410)$ in their $K\pi$ decay mode. We study the strong decay $K^*\to K\pi$  by employing a  nonzero total momentum $P$, where the $K^*(892)$ decay is kinematically facilitated on the lattice. For this purpose we simulate $K\pi$ scattering in $p$ waves with $P= \tfrac{2\pi}{L}e_z~,\ \tfrac{2\pi}{L}(e_x+e_y)$ and extract energy levels $E_n$. Each of the $E_n$ leads to the value of $\delta_{l=1}(s)$ at $s=E_n^2-P^2$ via the  generalized L\"uscher relations derived for the scattering of non-degenerate particles with nonzero $P$ in  \cite{Leskovec:2012gb}.  The resulting phases are combined with those obtained from our previous simulation at $P=0$ \cite{Lang:2012sv}. Finally we determine the resonance mass and width of $K^*(892)$, as well as the resonance mass of $K^*(1410)$  using a Breit-Wigner type formula.  

A challenging technical problem for simulating scattering of non-degenerate particles ($m_K\not =m_\pi$) at $P\not =0$ is that even and odd partial waves can in principle mix within one irreducible representation (irrep) of the discrete lattice group  \cite{Fu:2011xz,Leskovec:2012gb,Doring:2012eu,Gockeler:2012yj}. 
In the present simulation we use only the irreducible representations ($E$, $B_2$ and $B_3$) where the $p$ wave does not mix with the $s$ wave \cite{Leskovec:2012gb}; this is extremely important for a reliable extraction of the $p$ wave since the $s$ wave is non-negligible in the whole energy region. In fact 
the  $p$ wave can mix with the $d$ wave for the irreducible representations we consider \cite{Fu:2011xz,Leskovec:2012gb}, which is an artifact of the reduced discrete symmetry group on the lattice and does not appear in infinite continuum space-time.  The $d$-wave phase shift is experimentally known to be negligible up to the $d$-wave resonance $K_2^*(1430)$ \cite{Estabrooks:1977xe,Aston:1987ir},  where the phase has a rapid increase by $\pi$, as is expected from a narrow resonance. In the present simulation we indeed observe  $p$-wave resonances $K^*$ as well as the $d$-wave resonance $K_2^*(1430)$ in the same irreducible representations.  This forces us to attempt a preliminary analysis of the  energy region around $K^*(1410)$ and $K_2^*(1430)$ using the generalized L\"uscher-type relation \cite{Leskovec:2012gb} that takes into account the mixing of both waves. 

The resonant $I=1/2$ channel with $K^*$ resonances is expected to have a behavior significantly different to the non-resonant $I=3/2$ channel, and we will verify this by simulating explicitly both isospin channels. 

The phase shift for $K\pi$ scattering was reliably extracted from experiment long time ago by Estabrooks {\it et al.} \cite{Estabrooks:1977xe} and Aston {\it et al.} \cite{Aston:1987ir},  briefly reviewed in \cite{Lang:2012sv}.

The $p$-wave phase shift for $K\pi$ scattering has been extracted from the lattice only in our previous simulation \cite{Lang:2012sv}, where the phase shifts for all four channels ($p$ and $s$-wave with $I=1/2,~3/2$ ) show qualitative agreement with experiment.  That simulation was for $P=0$  and led only to one phase shift point near the $K^*$ resonance,  which did not allow a determination of the $K^*$ width. There was another recent simulation aimed at $K^*$ \cite{Fu:2012tj}, but it extracted the $p$-wave phase shift at $P=2\pi/L$ from the irreducible representation $A_1$, which mixes $s$ and $p$ waves. Since the $s$-wave phase shift is sizable in the region of the $K^*(892)$ resonance, we believe the extracted $p$-wave phase shift in \cite{Fu:2012tj} to be affected by sizable and unknown systematic uncertainties.    All other previous lattice simulations of $K\pi$ scattering studied the $s$-wave with $I=1/2,~3/2$ near threshold \cite{Sasaki:2009cz,Beane:2006gj,Fu:2011wc} and the resulting scattering lengths are compared in \cite{Lang:2012sv}.

The $K^*(892)\to K\pi$ coupling was already extracted once on the lattice using the so-called amplitude method, which is based on the  $\langle K\pi|K^*\rangle$ correlator  \cite{McNeile:2002fh}. This method assumes that the $K^*$ and $K\pi$ energies are equal, which is difficult to achieve in practice.

A review on the lattice studies of resonances is given in \cite{Mohler:2012nh}. The only resonance addressed by several lattice groups up to now is $\pi\pi\to \rho\to \pi\pi$ \cite{Dudek:2012xn,Pelissier:2012pi,Lang:2011mn,Feng:2010es,Aoki:2011yj,Aoki:2007rd}. Recently also the first simulation of charmed resonances \cite{Mohler:2012na}, as well as   $N^{-}$ \cite{Lang:2012db} and $\Delta$ \cite{Alexandrou:2013ata} resonances was performed. 

In phenomenological studies the $K^*$ resonance-pole emerged for example within the Roy-equation  approach \cite{Buettiker:2003pp,DescotesGenon:2006uk} and a unitarized version of the Chiral Perturbation Theory \cite{Oller:1998hw,Oller:1998zr,GomezNicola:2001as,Pelaez:2004xp,Guo:2011pa}. The latter approach has also been used to study the $m_{\pi,K}$ dependence \cite{Nebreda:2010wv} and the finite-volume effects \cite{Doring:2011nd,Doring:2012eu,Bernard:2010fp} in lattice simulations. 

Following the Introduction, we present our lattice setup in Section \ref{sec_lattice} and the interpolating fields (interp.) in Section \ref{sec_interpolators}, which are further detailed in the Appendix. Section \ref{sec_energies}  provides the energy levels for both isospin channels. Resulting $I=1/2$ phase shifts and $K^*$ resonances parameters are collected  in Section \ref{sec_half}, while  $I=3/2$ is considered in Section \ref{sec_threehalf}.

\section{Lattice setup}\label{sec_lattice}

Our simulation is based on one ensemble of gauge configurations with clover Wilson dynamical $u,\,d$ quarks and 
$u,\,d,\,s$ valence quarks ($m_s>m_u=m_d$), where the valence and the  dynamical $u/d$ quarks have the same mass.   The corresponding pion mass is $m_\pi= 266(4)~$MeV. The strange quark mass is fixed by $m_\phi$, which correspond to  $\kappa_s=0.12610$ and $c_{sw}=1$, rendering $m_K= 552(6)~$MeV. The parameters of the ensemble are shown in  Table \ref{gauge_configs}, while more details are given in \cite{Lang:2011mn,Lang:2012sv}. Due to the limited data on just a single ensemble, our determination of the lattice spacing $a$ reported in \cite{Lang:2011mn} results from taking a typical value of the Sommer parameter $r_0$. We note that the uncertainty associated with this choice might lead to small shift of all dimensionful quantities.  This ensemble has been generated by the authors of  \cite{Hasenfratz:2008ce,Hasenfratz:2008fg} to study re-weighting techniques. 

We have a rather small volume $V=16^3\times 32$ ($L\simeq 2~$fm), which enables us to use the powerful but costly full distillation method \cite{Peardon:2009gh}. This allows for the computation of all contractions for the correlation matrix of $\bar s u$ and $K\pi$ interpolators. 
The sea and valence quarks obey periodic boundary conditions in space. The periodic and anti-periodic valence propagators in time are combined into so-called $"P+A"$ propagators, which effectively extends the time direction to $2N_T=64$ \cite{Lang:2011mn,Lang:2012sv}.

\begin{table}[tb]
\begin{ruledtabular}
\begin{tabular}{ccccccc}
$N_L^3\times N_T$ & $\beta$ & $a$[fm] & $L$[fm] & \#cfgs & $m_\pi$[MeV]& $m_K$[MeV]\\ 
\hline
$16^3\times32$ & 7.1 & 0.1239(13) & 1.98 & 276 & 266(4) & 552(6)\\
\end{tabular}
\end{ruledtabular}
\caption{\label{gauge_configs} Parameters of the $N_f=2$ gauge configurations  \cite{Lang:2011mn,Lang:2012sv}.}\label{tab:lattice}
\end{table}

\begin{figure}[tb]
\begin{center}
\includegraphics*[width=0.35\textwidth,clip]{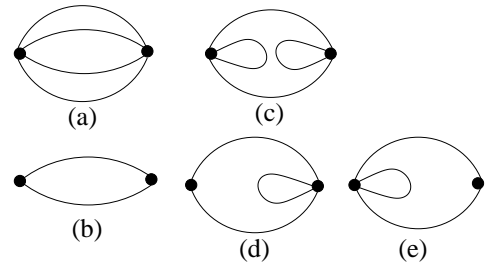}
\end{center}
\caption{Contractions for our correlators with $\bar su$ and $K\pi$ interpolators (\ref{O}), given in Appendix \ref{app_a}.  
For $I=3/2$ only (a) appears, while all are needed for $I=1/2$. 
The contractions (c,d)  need an all-to-all method like distillation.}\label{fig:contractions}
\end{figure}  

\section{Interpolating fields}\label{sec_interpolators}

The $K\pi$ physical system with momentum $P$ and $I=1/2$ or $I=3/2$ is created or annihilated with the interpolating fields listed
in the Appendix, having  the form
\begin{align}
\label{O}
{\cal O}_{I=1/2}^{\bar qq}&=\sum_{x} e^{iPx} \bar s(x) \hat \Gamma u(x)\FC\\
{\cal O}_{I=1/2}^{K\pi}&=\sum_j f_j[\sqrt{\tfrac{1}{3}}K^+(p_{Kj})\pi^0(p_{\pi j})\nonumber\FC\\
&\qquad\quad   +\sqrt{\tfrac{2}{3}}K^0(p_{Kj})\pi^+(p_{\pi j})]\nonumber\FC\\
 {\cal O}_{I=3/2}^{K\pi}&=\sum_j f_j ~K^+(p_{Kj})\pi^+(p_{\pi j}) ,\;\; p_{Kj}+p_{\pi j}=P\FD\nonumber
\end{align}
where $x$, $P$ and $p$ are three-vectors. 
Interpolators are constructed to transform according to irreducible representations as explained below and detailed in \cite{Leskovec:2012gb}. We employ five  ${\cal O}^{\bar qq}$ in each representation. In ${\cal O}^{K\pi}$ the momenta are projected separately for $K$ and $\pi$.  For each representation given below we use specific  linear combinations of momenta $p_{K,\pi}$ so that  ${\cal O}^{K\pi}$ transforms according to this irrep (irreducible representation): we use all possibilities with $p_K\leq \sqrt{3} \tfrac{2\pi}{L}$ and $p_\pi\leq \sqrt{3}\tfrac{2\pi}{L}$ according to \cite{Leskovec:2012gb}. 

In order to facilitate the $K^*\to K\pi$ decay kinematically and to access further values of $s=E_n^2-P^2$, we implement interpolators (\ref{O}) with  non-zero total momenta $P$ (we considered also all  permutations of $P$ and all possible directions of polarizations at given $P$): 
\begin{align}
\label{irreps}
P&\!=\!\tfrac{2\pi}{L}e_z:\,           &C_{4v}\,, ~&\mathrm{irreps}\;E(e_{x,y}),~E(e_x\!\pm\! e_y),~ l=1,2\nonumber \\
P&\!=\!\tfrac{2\pi}{L}(e_x\!+\!e_y): &C_{2v}\,, ~&\mathrm{irreps}\;B_2,~ B_3,\qquad\qquad~\quad l=1,2\nonumber\\
P&\!=\!0:                         &O_h\,, ~     &\mathrm{irrep}\;T_1^-,\qquad\qquad\qquad\quad~  l=1\FD
\end{align}
The zero-momentum case, studied in \cite{Lang:2012sv}, is listed for completeness since we will combine all these results.   The analytic framework for  $p$-wave scattering using the first two  momenta is described in detail in \cite{Leskovec:2012gb}, together with the symmetry considerations, appropriate interpolating fields and extraction of the phase shifts, so we only briefly review the main steps here. 

The symmetry group of the  mesh viewed from the center-of-momentum (CMF) frame of the $K\pi$ system is $C_{4v}$ for $P=\tfrac{2\pi}{L}e_z$ and $C_{2v}$ for $P=\tfrac{2\pi}{L}(e_x+e_y)$. These groups do not contain the inversion as an element, which  in turn implies that even and odd partial waves can in principle mix within the same irreducible representation. For extraction of $\delta_{l=1}$ a particularly disturbing mixing is the one with $\delta_{l=0}$, since $\delta_{0}(s)$ is known to be non-negligible in the whole energy region of interest. Fortunately $\delta_1$ does not mix with  $\delta_0$ in the irreducible representations $E,~B_2,~B_3$ (see equation (\ref{irreps})), so we can use these. In fact we employ two  representations of the two-dimensional $E$: $E(e_{x,y})$ with basis vectors along axis $(e_x,e_y)$ and $E(e_x\pm e_y)$ with basis vectors along the diagonal $(e_x+e_y,e_x-e_y)$. 

Each of the five representations  $B_2$, $B_3$, $E(e_{x,y})$, $E(e_x\pm e_y)$, and $T_1^-$ will lead to energy levels $E_n$, values of $s=E_n^2-P^2$ and scattering phases $\delta(s)$.  

The quarks in (\ref{O}) are smeared with Laplacian Heaviside smearing \cite{Peardon:2009gh}
\begin{equation}\label{smearing}
q_s(n)\equiv \sum_{k=1}^{N_v} v^{(k)}(n) v^{(k)\dagger}(n') ~q(n') \FC
\end{equation}
where $v^{(k)}$ are the eigenvectors of the 3D lattice Laplacian $\quad \nabla^2\,v^{(k)}=\lambda_{\nabla^2}^{(k)} \,v^{(k)}$ and $n$ and $n'$ are brief for the space and color indices.
We choose  $N_v=96$  for ${\cal O}^{\bar qq}$ and $N_v=64$ for the more costly ${\cal O}^{K\pi}$. 
This allows the calculation of all contractions  in Fig. \ref{fig:contractions}  according to the full distillation method \cite{Peardon:2009gh}.  

\section{Energy levels for $I=1/2,~3/2$ }\label{sec_energies}

The energy spectrum $E_n$ is extracted from the correlation matrix 
\begin{equation}
C_{ij}(t)=\tfrac{1}{N_T}\sum_{t_n} \langle {\cal O}_i^\dagger (t_n+t) |{\cal O}_j(t_n)\rangle\FC
\end{equation}
averaged over all initial times $t_n$. All needed Wick contractions, shown in  Fig. \ref{fig:contractions}, are evaluated.  We apply the generalized eigenvalue problem $C(t)u_n(t)=\lambda_n(t)C(t_0)u_n(t)$ \cite{Michael:1985ne,Luscher:1985dn,Luscher:1990ck,Blossier:2009kd}. The resulting eigenvalues $\lambda_n(t)\to \E^{-E_n (t-t_0)}$ give the effective energies $E_n^{eff}(t)\equiv \log [\lambda_n(t)/\lambda_n(t+1)]\to E_n$  and the eigenvectors are the fingerprints of the energy eigenstates. 

\begin{figure*}[tb]
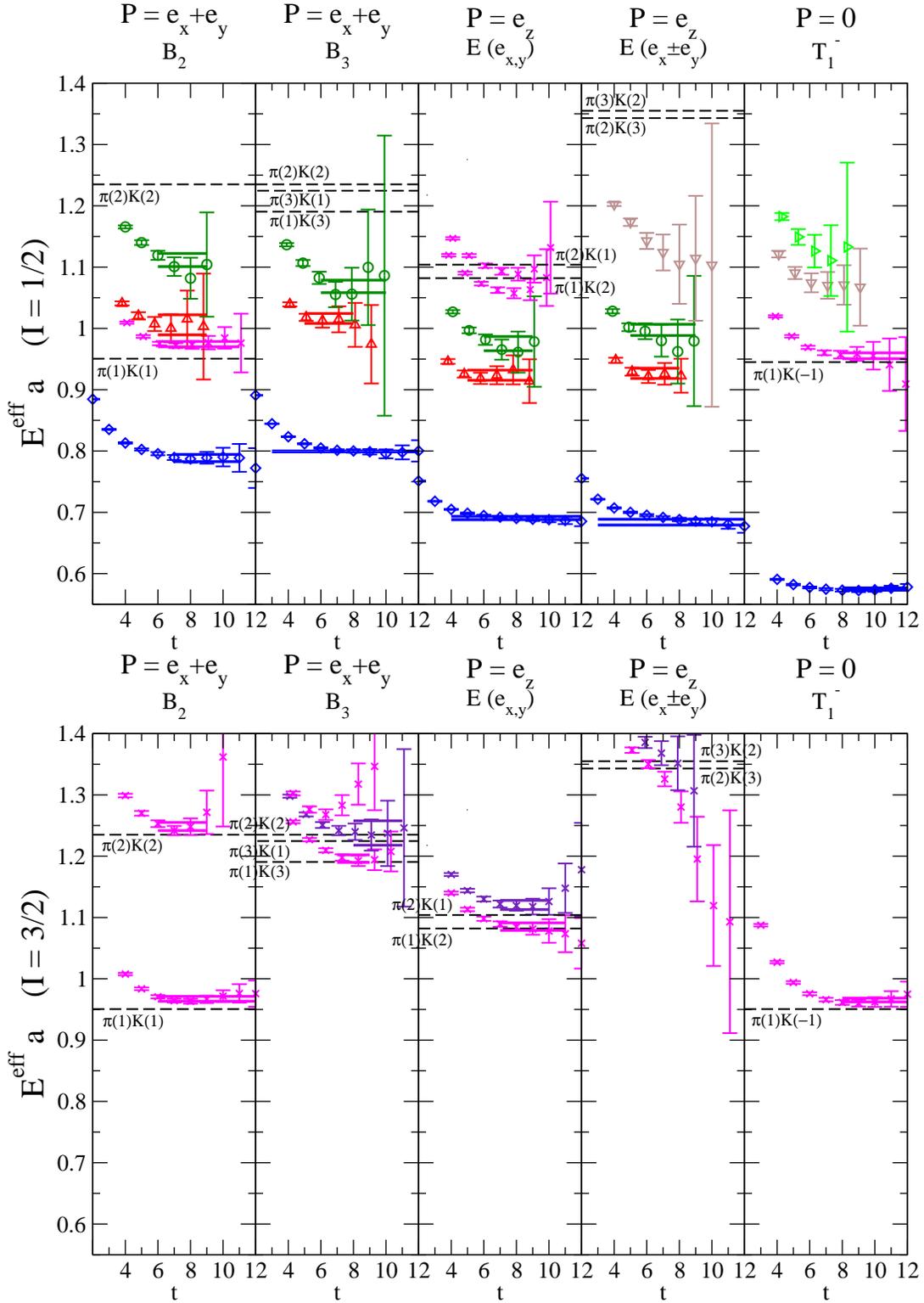

\begin{center}
\includegraphics*[width=0.8\textwidth,clip]{isospin_1-2.eps}
\includegraphics*[width=0.8\textwidth,clip]{isospin_3-2.eps} 
\end{center}
\caption{ Effective energies $E_n^{eff}(t)\,a$ of the $K\pi$ system  with $I=1/2,~3/2$ at total momentum $P\not =0$ and $P=0$ (in lattice units $a^{-1}\simeq 1.59$ GeV).  The dashed lines indicate energy vales $E_{n.i.}$  for the non-interacting scattering levels in the notation $K(n_K^2)\pi(n_\pi^2)$ (see equation (\ref{E_ni})).}\label{fig:spectrum}
\end{figure*}

A few lowest effective energies $E_n^{eff}(t)$ for the $I=1/2,~3/2$ states are shown  in Fig. \ref{fig:spectrum}.  The energies for the levels with reliable plateaus (marked by horizontal fits) are provided in  Tables \ref{tab:spectrum_half} and \ref{tab:spectrum_threehalf}. All error bars are determined by single elimination jackknife.

\begin{figure*}[htb]
\begin{center}
\includegraphics*[width=0.65\textwidth,clip]{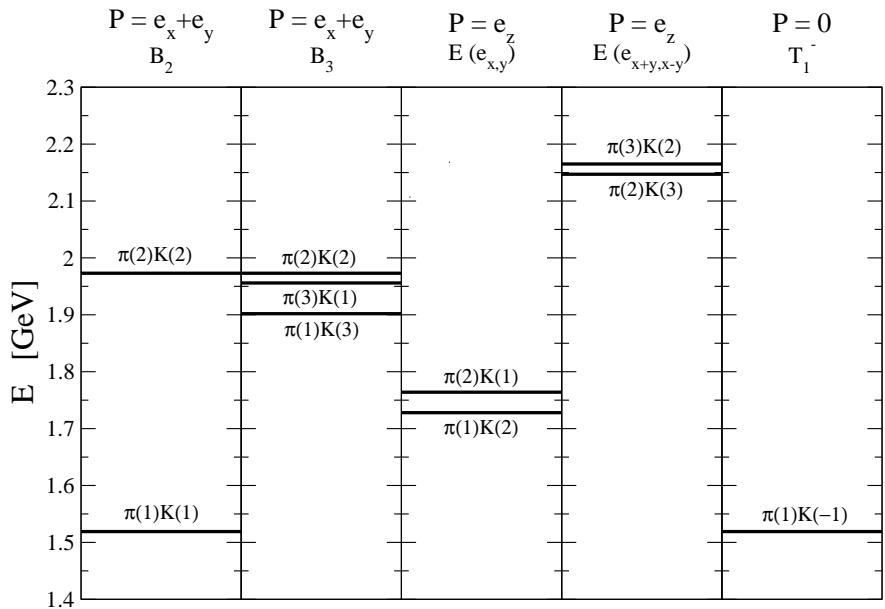} 
\end{center}
\caption{Energies for non-interacting $K\pi$ scattering states (\ref{E_ni}) on our lattice 
($m_\pi$ and $m_K$ are given in Table \ref{gauge_configs}).  The values correspond to the dashed lines in Fig. \ref{fig:spectrum}, now shown in GeV units.
}
\label{fig:sqrts-non-interact}
\end{figure*}

The energy spectrum of $K\pi$ scattering is discrete on a finite lattice. In absence of interaction, the scattering levels $K(n_K)\pi(n_\pi)$ are just sums of energies of the individual particles with
\begin{align}
\label{E_ni}
&E_{n.i.}({K(n_K)\pi(n_\pi)})=\sqrt{m_K^2+p_K^2}+\sqrt{m_\pi^2+p_\pi^2}\FC\nonumber \\
&\ p_{\pi}=\tfrac{2\pi}{L}n_{\pi}\FC\quad p_{K}=\tfrac{2\pi}{L}n_{K}\FC\quad n_{\pi,K}\in N^3\FC
\end{align}
which are given by the dashed lines  in Fig. \ref{fig:spectrum}.  In Fig. \ref{fig:sqrts-non-interact} we show the corresponding values of 
$E_{n.i.}$. 
This plot demonstrates the difficulty to achieve energy values for a scattering level in the region of the $K^*(892)$ resonance for a typical lattice size.
Note that the lowest eneregy is reached for $K(e_x)\pi(e_y)$ which appears in the $B_2$ representation.

In the interacting case the lattice energies $E_n$ get shifted with respect to $E_{n.i.}$. A small shift $\Delta E=E_n-E_{n.i.}$ 
corresponds to a small phase shift $\delta(s)$ at $s=E_n^2-P^2$ (modulo multiples of $\pi$). 
The $I=3/2$ scattering in Fig. \ref{fig:spectrum} is a typical example: all levels are near $E_{n.i.}$ and a small positive $\Delta E$ is related to a small and negative $\delta_1$ in this repulsive channel.  The  scattering levels $K(e_x)\pi(e_y)$ in $B_2$ and  $K(e_z)\pi(-e_z)$ in $T_1^-$ are also clearly visible in the $I=1/2$ channel; we also observe higher $I=1/2$ scattering  levels but  some are not included in the analysis and the plots due to less reliable plateaus\footnote{Higher $I=1/2$ scattering levels have less reliable plateaus than  $I=3/2$ ones due  to the contraction (c) in Fig. \ref{fig:contractions}. They also constitute already the $4^{th}$  level or higher for $I=1/2$.}. 

Narrow resonances  lead to levels in addition to the scattering levels $K(n_K)\pi(n_\pi)$. Such extra levels are indeed observed in Fig. \ref{fig:spectrum} for the resonant $I=1/2$  scattering. The $K^*(892)$ is narrow in experiment and even narrower on our lattice (due to smaller phase space at $m_\pi=266~$MeV), and is responsible for the lowest level with $I=1/2$ in all irreps. We find that for all irreps except $B_2$, the lowest energy is insensitive as to whether ${\cal O}^{K\pi}$ is included in the correlation matrix or not; this is expected since the scattering levels $K\pi$ have a high energy $E_{n.i.}$ (\ref{E_ni}) corresponding to $\sqrt{s}\!\gg\! m_{K^*(892)}$ (Fig. \ref{fig:sqrts-non-interact}) in these irreps, so they influence the ground state  only weakly.  
In the case of irrep $B_2$, which has the $K\pi$ scattering state  at lowest $\sqrt{s}$ in Fig. \ref{fig:sqrts-non-interact}, the ground state is slightly (but still within the error on $E^{eff}$) shifted down when ${\cal O}^{K\pi}$ is included in the basis.

There is one additional level in irreps $B_2,~B_3,~E$ near $\sqrt{s}\simeq 1.4~$GeV which we attribute  to the $K^*(1410)$ resonance. In fact, we find it puzzling that there is no additional level\footnote{In $T_1^-$, the level 2 at $E_2\simeq 1.5~$GeV  corresponds to $K(1)\pi(-1)$, while the next level comes only at $E_3\simeq 1.7~$GeV rendering $\delta\simeq 90^\circ$ \cite{Lang:2012sv}.  } 
near $\sqrt{s}\simeq m_{K^*(1410)}$ in $T_1^-$ \cite{Lang:2012sv} and we prompt future lattice simulations with $K\pi$ interpolators to shed  light on this point. 

\begin{figure}[htb]
\begin{center}
\includegraphics*[width=0.4\textwidth,clip]{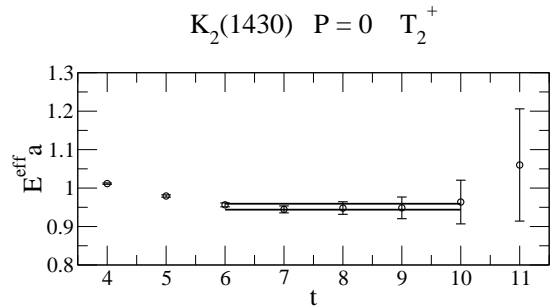} 
\end{center}
\caption{ Effective energy from quark-antiquark interpolators in the  $T_2^+$ irrep of $O_h$. The level is related to $K_2^*(1430)$ resonance in $d$-wave scattering of $K\pi$. We obtain $E\,a=0.9515(77)$.}\label{fig:T2}
\end{figure}

Note that there is another extra level near $\sqrt{s}\simeq 1.4~$GeV for irreps $B_2$,$B_3$, $E$ with $P\not =0$ and we attribute this to the resonance $K_2^*(1430)$. As we mentioned, the  $p$-wave scattering  mixes with $d$-wave scattering in  irreps $B_2$,$B_3$, $E$ and it is interesting that we indeed observe this mixing \cite{Fu:2011xz,Leskovec:2012gb}. Our interpretation of this level is supported by the fact that we find $K_2^*(1430)$ for $P=0$ at similar $\sqrt{s}$ in Fig. \ref{fig:T2}. In this case we employ $T_2^+$ of $O_h$ which does not mix $l=1$ with $l=0,2$. We use a  $2\times 2$ correlation matrix with interpolators
\begin{align}
&\bar s |\epsilon_{ijk}| \gamma_j \nabla_k u \FC\nonumber
&\bar s |\epsilon_{ijk}| \gamma_t \gamma_j \nabla_k u \FC
\end{align}
for $T_2^+$.  

\begin{figure*}[htb]
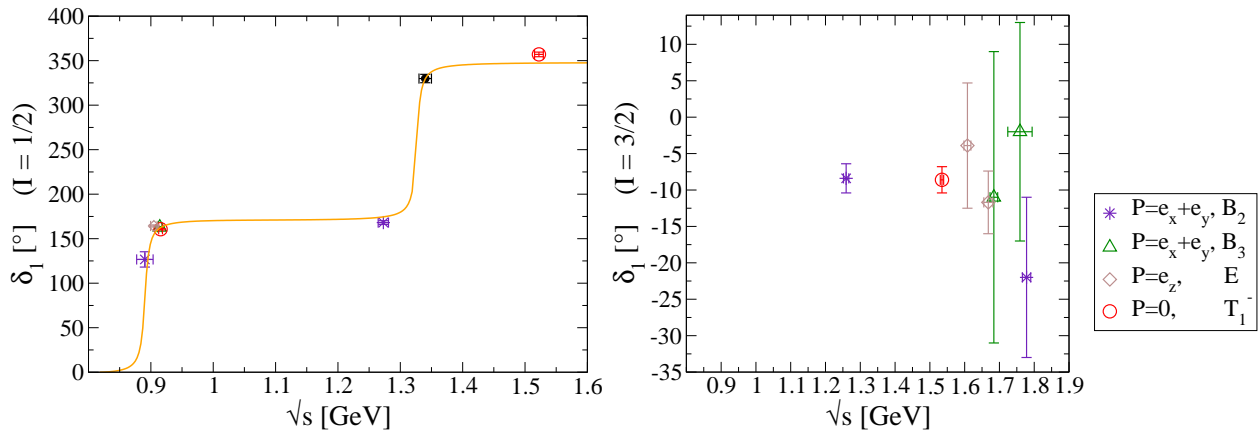

\begin{center}
\includegraphics*[width=0.44\textwidth,clip]{phase_1-2.eps}
\includegraphics*[width=0.48\textwidth,clip]{phase_3-2.eps}  
\end{center}
\caption{ The $p$-wave scattering phase shift $\delta_{l=1}$ as a function of $\sqrt{s}$ for $I=1/2$ and $I=3/2$. Different colors/symbols indicate results from different irreducible representations (\ref{irreps}), while the $\delta_1$ point obtained by taking into account  $\delta_{1,2}$ mixing is indicated by a black dot (left panel, around $\sqrt{s}\!=\!1.34~$GeV). Note that three points (circle, triangle and diamond) near $\sqrt{s}\simeq 0.91~$GeV are overlapping. The  line represents a fit over a pair of Breit-Wigner resonances (see equation (\ref{xy_sum})). 
}\label{fig:phases}
\end{figure*}

\section{Phase shifts and $K^*$ resonances in $I=1/2$ channel}\label{sec_half}

Each energy level $E_n$ from the previous section renders  a specific momentum $p^*=|p_K^{cmf}|=|p_\pi^{cmf}|$ of $\pi$ and $K$ in center-of-momentum frame via
\begin{align}
\label{q}
\sqrt{s}&=\sqrt{E_n^2-P^2}=\sqrt{m_\pi^2+p^{*2}}+\sqrt{m_K^2+p^{*2}}\FC\nonumber\\
q&\equiv\tfrac{L}{2\pi}p^*\FC
\end{align}
where $q$ is dimensionless.
Unlike in experiment, where $p^*$ is continuous due to $L=\infty$,
in our simulation we obtain only discrete values of $p^*$.

\subsubsection{Phase shift $\delta_1$ for $\sqrt{s}<1.3$ GeV and the $K^*(892)$ resonance }\label{sec:low_energy}

$K\pi$ scattering in $p$-wave is known to be elastic for $\sqrt{s}<1.3~$GeV experimentally \cite{Estabrooks:1977xe, Aston:1987ir}. In this region the elastic phase shift $\delta_{l=1}(s)$ at $s=E_n^2-P^2$ is reliably extracted for each value of  $q= p^*L/2\pi$ (or $E_n$). The relation $\delta_1(s)=\mathrm{atan} [\pi^{3/2}q/Z_{00}(1;q^2)]$ for $P=0$ was originally derived by L\"uscher in \cite{Luscher:1990ux,Luscher:1991cf}. For the case of $P\not =0$ the relevant L\"uscher-type relations were derived in \cite{Leskovec:2012gb}, where they are explicitly given by the equations (41), (42) and (56) for  the  irreducible representations $B_3$, $B_2$ and  $E$, respectively.
 These three relations neglect $\delta_{l=2}(s)$: this is a good approximation for the region $\sqrt{s}<1.3~$GeV since it is below $K_2^*(1430)$ \cite{Estabrooks:1977xe,Aston:1987ir}. The resulting $\delta_1$ is plotted  in Fig. \ref{fig:phases} and listed in Table \ref{tab:spectrum_half}.

The main  uncertainty in the resulting phases is the neglect of the exponential finite-volume corrections, which may not be completely negligible on our small volume and will have to be addressed in future simulations with larger $L$.  

The L\"uscher-like relations provide only $\tan(\delta_1)$, so the resulting phase is determined up to $\pm N\!\cdot\! 180^\circ$ and we choose $N$ such that the phase is rising with increasing $\sqrt{s}$ (as expected in a elastic resonant channel where $\delta$ increases by $180^\circ$ for each resonance). 

There are four phase shift points in the vicinity of $K^*(892)$ and a fast rise of the phase in a narrow region around $\sqrt{s}\simeq 0.89\;{\mathrm GeV}\simeq m_{K^*}$ is apparent. These four points will be used  for the  exploratory extraction of the $K^*(892)$. Note that phase shift points from $B_3$, $E$ and $T_1^-$, that  almost overlap in $\sqrt{s}$, overlap also in $\delta_1$; this is  a non-trivial check of the approach since L\"uscher's relations for these three irreducible representations have a different form \cite{Leskovec:2012gb,Lang:2012sv}.  

\begin{figure}[htb]
\begin{center}
\includegraphics*[width=0.48\textwidth,clip]{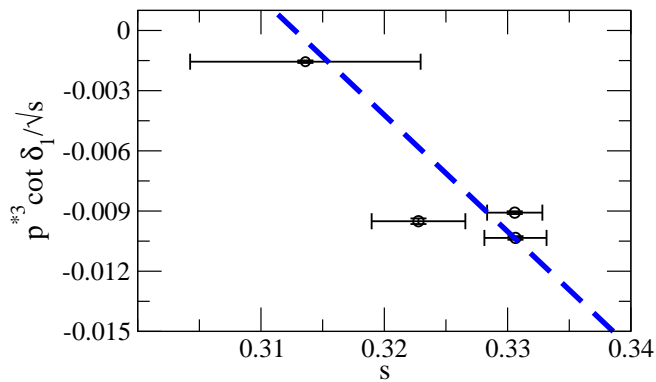} 
\end{center}
\caption{ The combination $\frac{(p^{*}a)^3}{\sqrt{sa^2}}\cot\delta_1(s)$ as a function of $sa^2$ in the vicinity of a narrow $K^*(892)$ resonance. The dependence is expected to be linear (\ref{xy})  for a  Breit-Wigner resonance and the  linear fit leads to $m_{K^*}$ and the coupling $g$ or $\Gamma[K^*\to K\pi]$ (\ref{amplitude}). 
}\label{fig:xy_Kstar}
\end{figure}

The four phase shift points with $\sqrt{s}$ near the narrow $K^*(892)$ are expected to be well described by the Breit-Wigner form   
\begin{align}
\label{amplitude}
T_l(s)&=\frac{\sqrt{s}\,\Gamma(s)}{m^2_{K^*}-s-\I \sqrt{s}
\Gamma(s)}\nonumber\\
&=
\frac{e^{2\I\delta_l(s)}-1}{2\I}=\frac{1}{\cot \delta_l(s)-\I}\FC\nonumber\\
\Gamma(s)&=\Gamma[K^*\to K\pi]=\frac{g^2}{6\pi}\frac{p^{*3}}{s}\FD
\end{align}
where the $K^*\to K\pi$ width $\Gamma$ is parametrized in terms of the phase space and the $K^*(892)\to K\pi$ coupling $g$. The  phase space is smaller for $m_\pi=266$ MeV than for $m_\pi^{phy}$, while the coupling $g$ is expected to be only mildly dependent on $m_\pi$, as explicitly verified within unitarized ChPT in \cite{Nebreda:2010wv}. So our main result will not be the width  but rather the coupling $g$, that will be compared to the experiment. 
The Breit-Wigner relation (\ref{amplitude}) can be rewritten in the form 
\begin{equation}
\label{xy}
\frac{p^{*3}}{\sqrt{s}}\cot\delta_1(s)=\frac{6\pi}{g^2}(m_{K^*}^2-s)
\end{equation}
and the values of the left-hand side are provided in Fig. \ref{fig:xy_Kstar} and Table \ref{tab:spectrum_half}. The linear fit in $s$ over the four phase shift points leads to $g$ and $m_{K^*}$ in Table \ref{tab:g} and these agree well with $m_{K^*}$ and $g$ derived from the experiment. Our results apply for  $m_{\pi,K}$ on our lattice,  but $m_{\pi}$ dependence of $g$ was shown to be very mild $g(m_\pi\!\!=\!\!266~\mathrm{MeV})/g(m_\pi^{phy})\simeq 1.03$ within unitarized ChPT, while its $m_K$ dependence is completely negligible \cite{Nebreda:2010wv}.   

This result can also be compared to $g=\sqrt{6\pi}~\bar g=\sqrt{6\pi}~1.44=6.25$ as obtained in the simulation \cite{McNeile:2002fh} using the amplitude method and assuming that the $K^*$ and $K\pi$ lattice energies are equal.

\begin{table}[h]
\begin{ruledtabular}
\begin{tabular}{c| c c | c c }
& $m_{K^*(892)}$ & $g_{K^*(892)}$ & $m_{K^*(1410)}$ & $g_{K^*(1410)}$  \\
& [MeV]       & [no unit] & [GeV] & [no unit] \\
  \hline
lat &  $ 891 \pm 14 $  & $ 5.7\pm 1.6 $ & $1.33 \pm 0.02$  & input  \\
exp &  $891.66 \pm 0.26$  &  $5.72 \pm 0.06 $& $1.414 \pm 0.0015$  &  $1.59\pm 0.03$   \\
\end{tabular}
\end{ruledtabular}
\caption{  The resulting resonance masses and  $K_i^*\to K\pi$ couplings $g$, which parametrize the width  $\Gamma[K_i^*\to K\pi]=(g_i^2p^{*3})/(6\pi s)$. The lattice results apply for our $m_\pi\simeq 266~$MeV and $m_K\simeq 552~$MeV, while the experimental couplings are derived from the observed $\Gamma[K_i^*\to K\pi]=Br[K_i^*\to K\pi]\Gamma_{K_i^*}$ and the values of $p^*$ and $s$ in  experiment.  }\label{tab:g}
\end{table}

\subsubsection{ The phase shift $\delta_1$ for $1.3<\sqrt{s}<1.6$ GeV and  $K^*(1410),\ K_2^*(1430)$ }

Unlike in the region $\sqrt{s}<1.3~$GeV, 
our exploratory extraction of the physics information from the energy levels in the region $\sqrt{s}>1.3~$GeV will inevitably be less reliable and based on certain approximations.

First of all, we  will assume that the $K\pi$ scattering in our simulation is elastic ($|1+2 \I T_l|=1$) up to $\sqrt{s}<1.6~$GeV, 
which is a strong approximation but indispensable for using L\"uscher's relations to extract the phase shift at present. In reality the $K\pi$ channel is coupled in this region to $K^*\pi$ and $K\rho$ channels, and experimentally  $Br[K^*(1410)\to K\pi] =6.6 \pm 1.3 \%$ while $Br[K^*(1680)\to K\pi]=38.7\pm 2.5 \%$. The treatment of such an inelastic problem is unfortunately  beyond the ability of current lattice simulations, although some practically very challenging approaches have been proposed analytically  \cite{Hansen:2012tf,Bernard:2010fp,Liu:2005kr,Doring:2011vk,Briceno:2012yi}.   In fact, we expect that the influence of $K^*\pi$ and $K\rho$ channels in our simulation is not significant, since we did not explicitly incorporate $K^*\pi$ and $K\rho$ interpolators\footnote{Similarly, most of previous simulations of meson resonances with $\bar qq$ interpolators assume that the scattering levels are not seen when they are not explicitly incorporated.}. 

The second complication stems from the fact that $d$-wave phase shift $\delta_2$ cannot be neglected around $\sqrt{s}\simeq m_{K_2^*(1430)}$ in  L\"uscher's relations. Therefore we derived the L\"uscher relations that contain $\delta_1$ as well as $\delta_2$ for irreps considered here: they are obtained from the so-called  determinant condition\footnote{For the original derivation of determinant condition see \cite{Luscher:1990ux,Rummukainen:1995vs,Kim:2005zzb}.} Eq. (28) in \cite{Leskovec:2012gb} by keeping non-zero $\delta_2$. For each irrep $B_3$, $B_2$, $E$ we get one (lengthy) phase shift equation (analog to Eqs. (41), (42), (56) in \cite{Leskovec:2012gb}), which depends on $q$ (see equation \ref{q}), $\delta_1(s)$ and $\delta_2(s)$. 

For a given level $E_n$ in a given irrep, we know $q$ (\ref{q}) and $s=E_n^2-P^2$, but one phase shift  equation alone cannot provide the values for two unknowns $\delta_1(s)$ and $\delta_2(s)$. Another level in another irrep unfortunately leads to two different unknowns $\delta_1(\tilde s)$ and $\delta_2(\tilde s)$, since this level in general corresponds to a different $\tilde s$ (see discussion in Section 3.1.3 of \cite{Leskovec:2012gb}). We overcome this serious difficulty by noting that the four levels with ''ID'' $K^*(1410)$  all have the invariant mass in a very narrow range of $\sqrt{s}=1.34 \pm 0.01~$GeV (see Table \ref{tab:spectrum_half}). By making a reasonable approximation that $s$ is the same for all four levels, we extract the unknown $\delta_1$ and $\delta_2$ by solving simultaneously two phase shift equations, namely  for\footnote{The levels $n=3$ in irreps $E(e_{x,y})$ and $E(e_x\pm e_y)$ occur at very similar $\sqrt{s}$, so they both lead to consistent $\delta_{1,2}$ via the same L\"uscher relation (56) in \cite{Leskovec:2012gb}. The errors on the resulting $\delta_{1,2}$ in (\ref{inelastic1},\ref{inelastic2}) are determined from the minimal and maximal values of $\sqrt{s}$ in the range $\sqrt{s}=1.34 \pm 0.01~$GeV.
}
\begin{align}
\label{inelastic1}
&\mathrm{level\ 3\ in\ irrep}\ B_2\mathrm{\ \&\ level\ 2\ in\ irrep}\ E~:\nonumber\\ 
&\qquad \sqrt{s}=1.34 \pm 0.01~\mathrm{GeV}\to\nonumber\\
&\delta_1=329.9^\circ \pm  4.4^\circ\quad \delta_2=89.6^\circ\pm 7.1^\circ~.
\end{align}
Then we extract $\delta_{1,2}$ from another pair of phase shift equations, corresponding to 
\begin{align}
\label{inelastic2}
&\mathrm{level\ 3\ in\ irrep}\ B_2\mathrm{\ \&\ level\ 2\ in\ irrep}\ B_3~:\nonumber\\ 
&\qquad \sqrt{s}=1.34 \pm 0.01~\mathrm{GeV}\to\nonumber\\
& \delta_1=329.8^\circ \pm 4.9^\circ \quad \delta_2=91.4^\circ\pm 6.2^\circ 
\end{align}
arriving at consistent results for the phases when compared to (\ref{inelastic1}), which indicates that our approximations are sensible. For the third pair of irreps, $B_3$ and $E$, we did not find a solution in the range of real $\delta_{1,2}$. The average $\delta_{1}$ from (\ref{inelastic1}) and  (\ref{inelastic2}) 
is provided for the corresponding levels in Table \ref{tab:spectrum_half} and by the black dot  in Fig. \ref{fig:phases}.

Finally we attempt an exploratory extraction of $K^*(1410)$ resonance parameters by fitting the resulting $\delta_1$ using a Breit-Wigner parametrization for two  resonances in the elastic region  (see equations (\ref{amplitude},\ref{xy}))
\begin{align}
\label{xy_sum}
&\frac{p^{*3}}{\sqrt{s}}\cot\delta_1(s)=\biggl[\sum_{K_i^*}\frac{g_{K_i^*}^2}{6\pi}\frac{1}{m_{K_i^*}^2-s}\biggr]^{-1}\FC\\
& K_i^*=K^*(892),~K^*(1410)\FD\nonumber
\end{align}
This satisfies $\cot\delta_1=0$ or $\delta=90^\circ \pm N\!\cdot\! 180^\circ$ at the position of each resonance $s=m_{K^*_i}^2$, while the relation of the phase (\ref{xy_sum}) to the amplitude $T_l$ (\ref{amplitude}) ensures the elasticity condition $|1+2\I T_l(s)|=1$. We fix   $m_{K^*(892)}$ and $g_{K^*(892)}$ in (\ref{xy_sum}) to the values  in Table \ref{tab:g}  extracted in this paper. The fit with two free parameters 
$g_{K^*(1410)}$ and $m_{K^*(1410)}$ is unfortunately not stable since there are only two phase shift points in the vicinity of $K^*(1410)$. 
Therefore we perform the Breit-Wigner\footnote{The resulting fit in Fig. \ref{fig:phases} indicates the weakness of the simple Breit-Wigner parametrization (see equations (\ref{xy},\ref{xy_sum})), where $\delta_1$ approaches $N\!\cdot\! 180^\circ$ to slowly at high  $\sqrt{s}$. An improved fit that incorporated damping of $p^{*3}$ in $\Gamma(s)$  at high energies was performed in the simulation of $\rho$ \cite{Dudek:2012xn}. Since such a fit contains additional free parameters, it is beyond our present analysis with only few phase shift points.}    fit (\ref{xy_sum})  over the few available points in Fig. \ref{fig:phases} 
by fixing $g_{K^*(1410)}$ to the value $g_{K^*(1410)}^{exp}=1.59\pm0.03$  derived from $\Gamma^{exp}[K^*(1410)\to K\pi]$. The resulting  resonance mass $m_{K^*(1410)}$ in Table  \ref{tab:g}\footnote{This is the only fit where the errors are not determined using the single-elimination jack-knife procedure due to the special treatment of the level near $\sqrt{s}\simeq 1.34~$GeV. The errors on $m_{K^*(1410)}$ follow from the variation of $\delta(s)$ and $g^{exp}_{K^*(1410)}$  in $1\sigma$ ranges. }
is  lower than in the experiment.

Let us point out again that our results for  $K^*(1410)$ resonance   rely on (i) the elasticity in the simulation, (ii) closeness of $s$ for certain levels (\ref{inelastic1},\ref{inelastic2}) and a Breit-Wigner fit (\ref{xy_sum}) over only a few values of the phase shift. Given these caveats, the reasonable agreement with experiment is satisfactory. 

\section{Phase shifts in $I=3/2$ channel} \label{sec_threehalf}

The $p$-wave scattering with $I=3/2$ was found to be elastic up to $\sqrt{s}<1.8~$GeV in experiment \cite{Estabrooks:1977xe}, while the $d$ wave with $I=3/2$ was found to be  negligible \cite{Estabrooks:1977xe}. So we extract $\delta_1$ assuming elasticity  and $\delta_2=0$, employing the same phase shift relations as for $I=1/2$ in Section \ref{sec:low_energy}. The resulting $\delta_1$ in  Fig. \ref{fig:spectrum} and Table \ref{tab:spectrum_threehalf} is  small and negative (or else consistent with zero\footnote{The phase is consistent with zero if $E$ is consistent with $E_{n.i.}$ (\ref{E_ni}). This is true for some of our higher lying levels, where it is challenging to accurately determine the energy shift $\Delta E=E-E_{n.i.}$. })  up to high $\sqrt{s}$, as expected in this repulsive channel with exotic isospin.  

\begin{table*}[tb]    
\begin{ruledtabular}
\begin{tabular}{ccc|ccccc|ccc|c}
$\tfrac{L}{2\pi}P$ & irrep & {\small level}  & $Ea$  & $\sqrt{s}~$[GeV] & $p^*a$  & $\delta_{1}~[^\circ]$ & $-~ \tfrac{\cot(\delta)(p^*a)^3}{\sqrt{sa^2}}$  & $t_0$ & interp. & fit      & ``ID'' \\
\hline
$e_x+e_y$&$B_2$&1& $0.7887(59)$ & $0.892(13)$ & $0.105(10)$ & $126.7(8.6)$ & $0.001557(64)$ & 2 & ${\cal O}^{\bar qq}_{1,3,4,5}\ {\cal O}^{K\pi}_{6,7}$ & 1exp$^u$: 7-11 & $K^{*}(892)$  \\                          
$e_x+e_y$&$B_2$&2& $0.9743(42)$ & $1.2749(83)$ & $0.2991(34)$ & $168.1(2.1)$ & $0.159(25)$ & 4 & ${\cal O}^{\bar qq}_{1,2,5}\ {\cal O}^{K\pi}_{6,7}$ & 1exp$^u$: 6-11 & $K(1)\pi(1)$  \\                             
$e_x+e_y$&$B_2$&3& $1.006(16)$ & $1.336(31)$ & $0.324(13)$ & $149.9(4.7)^{[*]}$ & $0.0328(14)^{[*]}$ & 4 & ${\cal O}^{\bar qq}_{1,2,5}\ {\cal O}^{K\pi}_{6,7}$ & 1exp$^u$: 6-9 & $K^*(1410)$ \\                      
$e_x+e_y$&$B_2$&4& $1.112(11)$ & $1.533(20)$ & $0.4000(74)$ &  &  & 4 & ${\cal O}^{\bar qq}_{1,2,5}\ {\cal O}^{K\pi}_{6,7}$ & 1exp$^u$: 6-9 & $K_{2}^*(1430)$ \\                                                     
\hline
$e_x+e_y$&$B_3$&1& $0.7994(16)$ & $0.9158(35)$ & $0.1226(24)$ & $162.8(0.7)$ & $0.010337(90)$ & 2 & ${\cal O}^{\bar qq}_{2,3,4,5}\ {\cal O}^{K\pi}_{6,7,8}$ & 2exp$^c$: 3-13 & $K^{*}(892)$  \\                      
$e_x+e_y$&$B_3$&2& $1.0164(81)$ & $1.356(15)$ & $0.3317(61)$ & $149.9(4.7)^{[*]}$ & $0.0328(14)^{[*]}$ & 4 & ${\cal O}^{\bar qq}_{2,3,4,5}\ {\cal O}^{K\pi}_{6,7,8}$ & 1exp$^u$: 5-8 & $K^{*}(1410)$  \\             
$e_x+e_y$&$B_3$&3& $1.073(15)$ & $1.462(28)$ & $0.373(11)$ &  &  & 4 & ${\cal O}^{\bar qq}_{2,3,4,5}\ {\cal O}^{K\pi}_{6,7,8}$ & 1exp$^u$: 6-10 & $K_{2}^{*}(1430)$  \\                                              
\hline
$e_z$&$E(e_{x,y})$&1& $0.6906(28)$ & $0.9048(53)$ & $0.1149(38)$ & $164.3(1.2)$ & $0.00951(14)$ & 2 & ${\cal O}^{\bar qq}_{1,2,3,4,5}\ {\cal O}^{K\pi}_{6,7}$ & 2exp$^c$: 4-15 & $K^{*}(892)$  \\                    
$e_z$&$E(e_{x,y})$&2& $0.9236(82)$ & $1.331(14)$ & $0.3220(58)$ & $149.9(4.7)^{[*]}$ & $0.0328(14)^{[*]}$ & 4 & ${\cal O}^{\bar qq}_{1,2,3,4,5}\ {\cal O}^{K\pi}_{6,7}$ & 1exp$^u$: 5-9 & $K^{*}(1410)$  \\          
$e_z$&$E(e_{x,y})$&3& $0.975(12)$ & $1.422(20)$ & $0.3575(78)$ &  &  & 4 & ${\cal O}^{\bar qq}_{1,2,3,4,5}\ {\cal O}^{K\pi}_{6,7}$ & 1exp$^u$: 6-9 & $K_{2}^{*}(1430)$  \\                                           
\hline
$e_z$&$E(e_{x \pm y})$&1& $0.6937(20)$ & $0.9107(39)$ & $0.1190(27)$ & $163.0(0.9)$ & $0.00966(10)$ & 2 & ${\cal O}^{\bar qq}_{1,2,3,4}\ {\cal O}^{K\pi}_{7}$ & 2exp$^c$: 3-14 & $K^{*}(892)$  \\                    
$e_z$&$E(e_{x \pm y})$&2& $0.9268(84)$ & $1.337(15)$ & $0.3242(59)$ & $149.9(4.7)^{[*]}$ & $0.0328(14)^{[*]}$ & 4 & ${\cal O}^{\bar qq}_{1,2,3,4,5}\ {\cal O}^{K\pi}_{6,7}$ & 1exp$^u$: 5-8 & $K^{*}(1410)$  \\      
$e_z$&$E(e_{x \pm y})$&3& $0.9977(92)$ & $1.461(16)$ & $0.3725(61)$ &  &  & 4 & ${\cal O}^{\bar qq}_{1,2,3,4,5}\ {\cal O}^{K\pi}_{6,7}$ & 1exp$^u$: 5-9 & $K_{2}^{*}(1430)$  \\                                      
\hline
$0$&$T_{1}^{-}$&1& $0.5749(19)$ & $0.9156(30)$ & $0.1225(21)$ & $160.6(0.7)$ & $0.00908(11)$ & 4 & ${\cal O}^{\bar qq}_{1,2,3}\ {\cal O}^{K\pi}_{6}$ & 1exp$^c$:8-16 & $K^{*}(892)$  \\                              
$0$&$T_{1}^{-}$&2& $0.9558(44)$ & $1.5223(70)$ & $0.3958(26)$ & $177.0(2.6)$ & $1.2(1.0)$ & 4 & ${\cal O}^{\bar qq}_{1,2,3}\ {\cal O}^{K\pi}_{6}$ & 1exp$^c$:8-12 & $K(1)\pi(-1)$                                    
\end{tabular}
\end{ruledtabular}
\caption{ Results for $K\pi$ scattering  in $p$-wave with $I=1/2$.    
Total momenta $P\not =0$ in different irreducible representations (\ref{irreps}) were considered in this work, while $P=0$ was simulated in  \cite{Lang:2012sv}. Here $E$ is energy in the lattice frame, inverse lattice spacing is $a^{-1}\simeq 1.592~$GeV, $\sqrt{s}=m_{K\pi}=\sqrt{E^2-P^2}$ is the $K\pi$ invariant mass and $p^*$  are the kaon/pion momenta in CMF (\ref{q}). The $\delta_1$ for levels  near $\sqrt{s}\simeq m_{K_2^*}$ is indicated by $^{[*]}$: it is an  average of (\ref{inelastic1}) and (\ref{inelastic2}), where the analysis takes into account the mixing of $p$-wave and $d$ wave at $P\not = 0$. The phases  extracted from L\"uscher-type relations are undetermined up to $\pm N\!\cdot\! 180^\circ$ and we choose $N$ such that absolute value of $\delta_1$ is rising with increasing $\sqrt{s}$.  Superscripts $c$ and $u$ in the fit indicate correlated and uncorrelated fits, respectively.  For easier identification ``ID'' indicates  the dominant Fock-component  according to our interpretation. }\label{tab:spectrum_half}

\vspace{0.3cm}

\begin{ruledtabular}
\begin{tabular}{ccc|ccccc|ccc|c}
$\tfrac{L}{2\pi}P$ & irrep & {\small level}  & $Ea$  & $\sqrt{s}~$[GeV] & $p^*a$  & $\delta~[^\circ]$ & $ -~ \tfrac{\cot(\delta)(p^*a)^3}{\sqrt{sa^2}}$  & $t_0$ & interp. & fit         & ``ID'' \\
\hline
$e_x+e_y$ & $B_2$ & 1 & $0.9674(39)$ & $1.2615(77)$ & $0.2935(32)$ & $-8.4(2.0)$ & $0.215(44)$ & 4 & ${\cal O}^{K\pi}_{6,7}$ & 1exp$^u$: 6-14 & $K(1)\pi(1)$  \\                                                   
$e_x+e_y$ & $B_2$ & 2 & $1.2484(66)$ & $1.781(12)$ & $0.4900(42)$ & $-22(11)$ & $0.26(13)$ & 4 & ${\cal O}^{K\pi}_{6,7}$ & 1exp$^u$: 6-9 & $K(2)\pi(2)$  \\                                                        
\hline
$e_x+e_y$ & $B_3$ & 1 & $1.1959(61)$ & $1.687(11)$ & $0.4564(40)$ & $-11(20)$ & $0.49(88)$ & 4 & ${\cal O}^{K\pi}_{6,7,8}$ & 1exp$^u$: 7-9 & $K(3)\pi(1)$  \\                                                      
$e_x+e_y$ & $B_3$ & 2 & $1.238(20)$ & $1.762(35)$ & $0.483(13)$ & $-2(15)$ & $3(41)$ & 4 & ${\cal O}^{K\pi}_{6,7,8}$ & 1exp$^u$: 8-11 & $K(1)\pi(3)$  \\                                                           
\hline
$e_z$ & $E(e_{x,y})$ & 1 & $1.0852(62)$ & $1.611(11)$ & $0.4288(39)$ & $-3.9(8.6)$ & $1.2(2.7)$ & 4 & ${\cal O}^{K\pi}_{6,7}$ & 1exp$^u$: 7-11 & $K(2)\pi(1)$  \\                                                  
$e_z$ & $E(e_{x,y})$ & 2 & $1.1204(74)$ & $1.671(13)$ & $0.4507(46)$ & $-11.7(4.3)$ & $0.42(15)$ & 4 & ${\cal O}^{K\pi}_{6,7}$ & 1exp$^u$: 7-10 & $K(1)\pi(2)$  \\                                                 
\hline
$0$ & $T_{1}^{-}$ & 1 & $ 0.9653(31)$ & $1.5374(49)$ & $0.4015(18)$ & $-8.6(1.8)$ & $0.443(91)$ & / & ${\cal O}^{K\pi}_{6}$ & 1exp$^u$:8-14 & $K(1)\pi(-1)$                                                        
\end{tabular}
\end{ruledtabular}
\caption{ Same as Table \ref{tab:spectrum_half} but for the $I=3/2$ channel.  }\label{tab:spectrum_threehalf}
\end{table*}
\section{Conclusions}\label{sec_conclusions}

We presented an exploratory study aimed to extract the masses and widths of $K^*(892)$ and $K^*(1410)$ resonances. For that purpose we simulated the $K\pi$ scattering in $p$ waves and extracted the $I=1/2,~3/2$ phase shifts $\delta_{l=1}$ shown in Fig. \ref{fig:phases}.   
The $K\pi$ system can have only discrete values of  the invariant mass  $s=\sqrt{E^2-P^2}$
due to the discretized momentum on the finite lattice. The  values in  Fig. \ref{fig:phases} were obtained by combining results from scattering with total momentum $P\not =0$ and $P=0$. 

The resonant $I=1/2$ and non-resonant $I=3/2$ channels show the expected differences in lattice spectrum and phase shift. All energy levels in the $I=3/2$ channel are near the expected $K\pi$ scattering states. In the $I=1/2$ channel we find  the $K\pi$ scattering states as well as additional energy levels near the resonances $K^*(892)$ and  $K^*(1410)$.  The $I=3/2$ phase shift is negative and small up to high energies, while the $I=1/2$ phase shows steep jumps at the $K^*(892)$ and $K^*(1410)$ resonances.

The  Breit-Wigner fit over the four phase shift points near $\sqrt{s}\simeq m_{K^*(892)}$ leads to the  $K^*(892)\to K\pi$ coupling (that parametrizes  $\Gamma_{K^*}$) and the $K^*(892)$ resonance mass in Table \ref{tab:g}. They agree with the experimental values within error. 
Our treatment of $K^*(892)$ is rigorous, since this region is elastic and 
the $p$-wave completely dominates the considered irreducible representations. The remaining  uncertainty is due to the finite volume and the lattice spacing,  which would have to be systematically addressed in future lattice simulations. 

The  $K^*(1410)$ resonance mass is extracted by fitting $\delta_1$ in Fig. \ref{fig:phases} for $\sqrt{s}<1.6~$GeV with two Breit-Wigner resonances (\ref{xy_sum}). In this fit the   $K^*(1410)\to K\pi$ coupling is fixed to the experimental value and the $K^*(892)$ values are fixed to our lattice results. The extracted  $K^*(1410)$ resonance mass in Table \ref{tab:g} is slightly lower than  in experiment.
We note that our analysis of the region near the $K^*(1410)$ resonance is inevitably less rigorous, as it neglects the possible presence of $K^*\pi$ and $K\rho$ channels, which might be a good approximation in our simulation without explicit $K^*\pi$ and $K\rho$ 
interpolators. An additional challenge in the region near $K^*(1410)$ comes from the mixing of $p$-wave and $d$-wave for $P\not =0$, which we take into account near the $K^*(1410)$ and $K_2^*(1430)$ resonances. 

This exploratory simulation of the $K^*(892)\to K\pi$ strong decay  is a first step toward treating 
the weak form factors $B\to K^*$ and $D\to K^*$ while  taking into account the $K^*\to K\pi$ decay. Some analytic ideas along these lines have already been proposed  in \cite{Bernard:2012bi}. This would be a challenging, but an important endeavor, since the resonant nature of $K^*$ is  limiting the precision at which phenomenologically important quantities are extracted from the lattice at present \cite{Liu:2011raa,Becirevic:2006nm}. 

\acknowledgments
We  thank Anna Hasenfratz for providing the
gauge configurations used for this work.  We would like to thank J.~Bulava, S.~Descotes-Genon, C.~Morningstar and  C.~Thomas for valuable discussions. The calculations were performed at Jozef Stefan Institute. This work is supported by the Slovenian Research Agency. Fermilab is operated by Fermi Research Alliance, LLC under Contract No. De-AC02-07CH11359 with the United States Department of Energy.

\clearpage
\begin{appendix}
\section{Interpolators for $K\pi$ scattering in $p$-wave with $I=1/2,~3/2$}\label{app_a}

Here we provide the  $I=1/2$ and $I=3/2$ interpolators ${\cal O}^{\bar qq}$ and ${\cal O}^{K\pi}$ with total momentum $P$ for  irreducible representations $B_2,~B_3,~E(e_{x,y}),~E(e_x+e_y)$, which contain $K\pi$ scattering with $l=1$ in continuum (see Section \ref{sec_interpolators}). These interpolators were proposed in \cite{Leskovec:2012gb}, where the correct transformation properties of these were demonstrated. The interpolators for irrep $T_1^-$ at $P=0$ were presented and simulated in \cite{Lang:2012sv}.

\subsection{$I=1/2$}
In each irrep we use five ${\cal O}_{op}^{\bar qq}$ ($op=1,..,5$), which are all constructed from vector currents $V_{i}^{op}(P)$ with polarization $i$
\begin{align}
V_{i}^{op=1}&={\textstyle\sum_{x}}\,\bar s(x)\,\gamma_i e^{iPx}\,u(x)\FC \\
V_{i}^{op=2}&={\textstyle\sum_{x}}\,\bar s(x)\,\gamma_t\gamma_i e^{iPx}\,u(x)\nonumber\FC \\
V_{i}^{op=3}&={\textstyle\sum_{x,j}}\,\bar s(x)\,
\overleftarrow{\nabla}_j\,\gamma_i\,\overrightarrow{\nabla}_j\,e^{iPx}u(x)\nonumber\FC \\
V_{i}^{op=4}&={\textstyle\sum_{x}}\,\bar s(x)\,
\tfrac{1}{2} \left[\overrightarrow{\nabla}_i -\overleftarrow{\nabla}_i\right]\,e^{iPx}u(x) 
\nonumber\FC \\
V_{i}^{op=5}&={\textstyle\sum_{x,j,k}} \,\epsilon_{ijk} 
\,\bar s(x)\gamma_j\gamma_5 \,\tfrac{1}{2}
\left[\overrightarrow{\nabla}_k  -\overleftarrow{\nabla}_k\right] 
\, e^{iPx}u(x)\FD\nonumber
\end{align}
The ${\cal O}^{K\pi}$ interpolators are linear combinations of $K(p_K)\pi(p_\pi)$ where momenta for $K$ and $\pi$ are separately projected  
\begin{align}
K^+(p_K)&=\sum_x e^{ip_K x}\bar s(x) \gamma_5 u(x)\FC \\
\pi^+(p_\pi)&=\sum_xe^{ip_\pi x}\bar d(x) \gamma_5 u(x)\FC\nonumber
\end{align}
and analogously for $K^0$ and $\pi^0$. 

In practice we simulated three permutations of direction $P=\tfrac{2\pi}{L}e_z$, and  three  of direction $P=\tfrac{2\pi}{L}(e_x+e_y)$, but present only interpolators for one choice. 

\subsubsection{Interpolators for irrep $B_2$ ($C_{2v}$) with $P=\tfrac{2\pi}{L}(e_x+e_y)$}
\begin{align}
{\cal O}^{\bar qq}_{op=1,..,5}=&V_x^{op}(e_x+e_y)-V_y^{op}(e_x+e_y)\FC \\
{\cal O}^{K\pi}_6=\sqrt{\tfrac{1}{3}}&\bigl[\pi^0(e_x)K^+(e_y)-\pi^0(e_y)K^+(e_x)\bigr]+\sqrt{\tfrac{2}{3}}\bigl[..\bigr]\nonumber\FC \\
{\cal O}^{K\pi}_7=\sqrt{\tfrac{1}{3}}&\bigl[\pi^0(e_x+e_z)K^+(e_y-e_z) \nonumber \\
-\pi^0(e_y&+e_z)K^+(e_x-e_z) +  \{ e_z\leftrightarrow -e_z\}\bigr]+\sqrt{\tfrac{2}{3}}\bigl[..\bigr]\FC\nonumber
\end{align}
where $[..]$ indicates another term with replacement $\pi^0\to \pi^+$ and $K^+\to K^0$.  Momenta $K$ and $\pi$ in ${\cal O}^{K\pi}$ are given in units of $2\pi/L$.

\subsubsection{Interpolators for irrep $B_3$ ($C_{2v}$) with $P=\tfrac{2\pi}{L}(e_x+e_y)$}
\begin{align}
{\cal O}^{\bar qq}_{op=1,..,5}=&V_z^{op}(e_x+e_y)\FC \\
{\cal O}^{K\pi}_6=\sqrt{\tfrac{1}{3}}&\bigl[\pi^0(e_x+e_y+e_z)K^+(-e_z)\nonumber\\-&\pi^0(e_x+e_y-e_z)K^+(e_z)\bigr]+\sqrt{\tfrac{2}{3}}\bigl[..\bigr]\nonumber\FC \\
{\cal O}^{K\pi}_7=\sqrt{\tfrac{1}{3}}&\bigl[K^+(e_x+e_y+e_z)\pi^0(-e_z)\nonumber\\-&K^+(e_x+e_y-e_z)\pi^0(e_z)\bigr]+\sqrt{\tfrac{2}{3}}\bigl[..\bigr]\nonumber\FC \\
{\cal O}^{K\pi}_8=\sqrt{\tfrac{1}{3}}&\bigl[\pi^0(e_x+e_z)K^+(e_y-e_z) \nonumber \\
+\pi^0(e_y&+e_z)K^+(e_x-e_z) -  \{ e_z\leftrightarrow -e_z\}\bigr]+\sqrt{\tfrac{2}{3}}\bigl[..\bigr]\FD\nonumber
\end{align}

\subsubsection{Interpolators for irrep $E~(e_{x,y})$ ($C_{4v}$) with $P=\tfrac{2\pi}{L}e_z$}
The two-dimensional basis is $(e_x,e_y)$ and we list interpolators for $e_x$ (but simulate also $e_y$):
\begin{align}
{\cal O}^{\bar qq}_{op=1,..,5}=&V_x^{op}(e_z)\FC \\
{\cal O}^{K\pi}_6=\sqrt{\tfrac{1}{3}}&\bigl[\pi^0(e_z+e_x)K^+(-e_x)\nonumber\\-&\pi^0(e_z-e_x)K^+(e_x)\bigr]+\sqrt{\tfrac{2}{3}}\bigl[..\bigr]\nonumber\FC \\
{\cal O}^{K\pi}_7=\sqrt{\tfrac{1}{3}}&\bigl[K^+(e_z+e_x)\pi^0(-e_x)\nonumber\\-&K^+(e_z-e_x)\pi^0(e_x)\bigr]+\sqrt{\tfrac{2}{3}}\bigl[..\bigr]\FD\nonumber 
\end{align}

\subsubsection{Interpolators for irrep $E~(e_{x}\pm e_y)$ ($C_{4v}$) with $P=\tfrac{2\pi}{L}e_z$}
The two-dimensional basis is $(e_x+e_y,e_x-e_y)$ and we list interpolators for $e_x-e_y$:
\begin{align}
{\cal O}^{\bar qq}_{op=1,..,5}=&V_x^{op}(e_z)-V_y^{op}(e_z)\FC \\
{\cal O}^{K\pi}_6=\sqrt{\tfrac{1}{3}}&\bigl[\pi^0(e_z+e_x-e_y)K^+(-e_x+e_y)\nonumber\\-&\pi^0(e_z-e_x+e_y)K^+(e_x-e_y)\bigr]+\sqrt{\tfrac{2}{3}}\bigl[..\bigr]\nonumber\FC \\
{\cal O}^{K\pi}_7=\sqrt{\tfrac{1}{3}}&\bigl[K^+(e_z+e_x-e_y)\pi^0(-e_x+e_y)\nonumber\\-&K^+(e_z-e_x+e_y)\pi^0(e_x-e_y)\bigr]+\sqrt{\tfrac{2}{3}}\bigl[..\bigr]\FD\nonumber 
\end{align}

\subsection{$I=3/2$}
There are no quark-antiquark interpolators with $I=3/2$, so we incorporated only $K^+\pi^+$ interpolators. We employ all ${\cal O}^{K\pi}$ listed for $I=1/2$ with the obvious replacement $\sqrt{1/3}~\pi^0K^++\sqrt{2/3}~\pi^+K^0\longrightarrow K^+\pi^+$. 

\end{appendix}

\bibliographystyle{h-physrev4}
\bibliography{Lgt}

\begin{thebibliography}{10}

\bibitem{Liu:2011raa}
Z.~Liu {\em et~al.},
\newblock [arXiv:1101.2726].

\bibitem{Becirevic:2006nm}
D.~Becirevic, V.~Lubicz and F.~Mescia,
\newblock Nucl.Phys. {\bf B769}, 31 (2007), [arXiv:hep-ph/0611295].

\bibitem{Leskovec:2012gb}
L.~Leskovec and S.~Prelovsek,
\newblock Phys. Rev. D {\bf 85}, 114507 (2012), [arXiv:1202.2145].

\bibitem{Lang:2012sv}
C.~B. Lang, L.~Leskovec, D.~Mohler and S.~Prelovsek,
\newblock Phys.Rev. {\bf D86}, 054508 (2012), [arXiv:1207.3204].

\bibitem{Fu:2011xz}
Z.~Fu,
\newblock Phys. Rev. {\bf D85}, 014506 (2012), [arXiv:1110.0319].

\bibitem{Doring:2012eu}
M.~D{\"o}ring, U.~Meissner, E.~Oset and A.~Rusetsky,
\newblock Eur.Phys.J. {\bf A48}, 114 (2012), [arXiv:1205.4838].

\bibitem{Gockeler:2012yj}
M.~G{\"o}ckeler {\em et~al.},
\newblock Phys.Rev. {\bf D86}, 094513 (2012), [arXiv:1206.4141].

\bibitem{Estabrooks:1977xe}
P.~Estabrooks {\em et~al.},
\newblock Nucl. Phys. {\bf B133}, 490 (1978).

\bibitem{Aston:1987ir}
D.~Aston {\em et~al.},
\newblock Nucl.Phys. {\bf B296}, 493 (1988).

\bibitem{Fu:2012tj}
Z.~Fu and K.~Fu,
\newblock Phys.Rev. {\bf D86}, 094507 (2012), [arXiv:1209.0350].

\bibitem{Sasaki:2009cz}
K.~Sasaki, N.~Ishizuka, T.~Yamazaki and M.~Oka,
\newblock PoS {\bf LAT2009}, 098 (2009), [arXiv:0911.0228].

\bibitem{Beane:2006gj}
S.~R. Beane {\em et~al.},
\newblock Phys. Rev. D {\bf 74}, 114503 (2006), [arXiv:hep-lat/0607036].

\bibitem{Fu:2011wc}
Z.~Fu,
\newblock Phys. Rev. D {\bf 85}, 074501 (2012), [arXiv:1110.1422].

\bibitem{McNeile:2002fh}
UKQCD Collaboration, C.~McNeile and C.~Michael,
\newblock Phys.Lett. {\bf B556}, 177 (2003), [arXiv:hep-lat/0212020].

\bibitem{Mohler:2012nh}
D.~Mohler,
\newblock PoS {\bf LATTICE2012}, 003 (2012), [arXiv:1211.6163].

\bibitem{Dudek:2012xn}
J.~J. Dudek, R.~G. Edwards and C.~E. Thomas,
\newblock Phys.Rev. {\bf D87}, 034505 (2013), [arXiv:1212.0830].

\bibitem{Pelissier:2012pi}
C.~Pelissier and A.~Alexandru,
\newblock Phys.Rev. {\bf D87}, 014503 (2013), [arXiv:1211.0092].

\bibitem{Lang:2011mn}
C.~B. Lang, D.~Mohler, S.~Prelovsek and M.~Vidmar,
\newblock Phys. Rev. D {\bf 84}, 054503 (2011), [arXiv:1105.5636].

\bibitem{Feng:2010es}
X.~Feng, K.~Jansen and D.~B. Renner,
\newblock Phys. Rev. D {\bf 83}, 094505 (2011), [arXiv:1011.5288].

\bibitem{Aoki:2011yj}
CS, S.~Aoki {\em et~al.},
\newblock Phys. Rev. D {\bf 84}, 094505 (2011), [arXiv:1106.5365].

\bibitem{Aoki:2007rd}
CP-PACS, S.~Aoki {\em et~al.},
\newblock Phys. Rev. D {\bf 76}, 094506 (2007), [arXiv:0708.3705].

\bibitem{Mohler:2012na}
D.~Mohler, S.~Prelovsek and R.~Woloshyn,
\newblock Phys.Rev. {\bf D87}, 034501 (2013), [arXiv:1208.4059].

\bibitem{Lang:2012db}
C.~B. Lang and V.~Verduci,
\newblock Phys. Rev. {\bf D87}, 054502 (2013), [arXiv:1212.5055].

\bibitem{Alexandrou:2013ata}
C.~Alexandrou, J.~Negele, M.~Petschlies, A.~Strelchenko and A.~Tsapalis,
\newblock [arXiv:1305.6081].

\bibitem{Buettiker:2003pp}
P.~Buettiker, S.~Descotes-Genon and B.~Moussallam,
\newblock Eur.Phys.J. {\bf C33}, 409 (2004), [arXiv:hep-ph/0310283].

\bibitem{DescotesGenon:2006uk}
S.~Descotes-Genon and B.~Moussallam,
\newblock Eur. Phys. J. {\bf C48}, 553 (2006), [arXiv:hep-ph/0607133].

\bibitem{Oller:1998hw}
J.~Oller, E.~Oset and J.~Pelaez,
\newblock Phys.Rev. {\bf D59}, 074001 (1999), [arXiv:hep-ph/9804209].

\bibitem{Oller:1998zr}
J.~Oller and E.~Oset,
\newblock Phys.Rev. {\bf D60}, 074023 (1999), [arXiv:hep-ph/9809337].

\bibitem{GomezNicola:2001as}
A.~Gomez~Nicola and J.~Pelaez,
\newblock Phys.Rev. {\bf D65}, 054009 (2002), [arXiv:hep-ph/0109056].

\bibitem{Pelaez:2004xp}
J.~Pelaez,
\newblock Mod.Phys.Lett. {\bf A19}, 2879 (2004), [arXiv:hep-ph/0411107].

\bibitem{Guo:2011pa}
Z.-H. Guo and J.~Oller,
\newblock Phys. Rev. D {\bf 84}, 034005 (2011), [arXiv:1104.2849].

\bibitem{Nebreda:2010wv}
J.~Nebreda and J.~Pelaez,
\newblock Phys. Rev. D {\bf 81}, 054035 (2010), [arXiv:1001.5237].

\bibitem{Doring:2011nd}
M.~D{\"o}ring and U.-G. Mei{\ss}ner,
\newblock JHEP {\bf 1201}, 009 (2012), [arXiv:1111.0616].

\bibitem{Bernard:2010fp}
V.~Bernard, M.~Lage, U.~G. Meissner and A.~Rusetsky,
\newblock JHEP {\bf 01}, 019 (2011), [arXiv:1010.6018].

\bibitem{Hasenfratz:2008ce}
A.~Hasenfratz, R.~Hoffmann and S.~Schaefer,
\newblock Phys. Rev. D {\bf 78}, 054511 (2008), [arXiv:0806.4586].

\bibitem{Hasenfratz:2008fg}
A.~Hasenfratz, R.~Hoffmann and S.~Schaefer,
\newblock Phys. Rev. D {\bf 78}, 014515 (2008), [arXiv:0805.2369].

\bibitem{Peardon:2009gh}
Hadron Spectrum Collaboration, M.~Peardon {\em et~al.},
\newblock Phys. Rev. D {\bf 80}, 054506 (2009), [arXiv:0905.2160].

\bibitem{Michael:1985ne}
C.~Michael,
\newblock Nucl. Phys. B {\bf 259}, 58 (1985).

\bibitem{Luscher:1985dn}
M.~L{\"u}scher,
\newblock Commun. Math. Phys. {\bf 104}, 177 (1986).

\bibitem{Luscher:1990ck}
M.~L{\"u}scher and U.~Wolff,
\newblock Nucl. Phys. B {\bf 339}, 222 (1990).

\bibitem{Blossier:2009kd}
B.~Blossier, M.~DellaMorte, G.~von Hippel, T.~Mendes and R.~Sommer,
\newblock JHEP {\bf 0904}, 094 (2009), [arXiv:0902.1265].

\bibitem{Luscher:1990ux}
M.~L{\"u}scher,
\newblock Nucl. Phys. B {\bf 354}, 531 (1991).

\bibitem{Luscher:1991cf}
M.~L{\"u}scher,
\newblock Nucl. Phys. B {\bf 364}, 237 (1991).

\bibitem{Hansen:2012tf}
M.~T. Hansen and S.~R. Sharpe,
\newblock Phys.Rev. {\bf D86}, 016007 (2012), [arXiv:1204.0826].

\bibitem{Liu:2005kr}
C.~Liu, X.~Feng and S.~He,
\newblock Int.J.Mod.Phys. {\bf A21}, 847 (2006), [arXiv:hep-lat/0508022].

\bibitem{Doring:2011vk}
M.~D{\"o}ring, U.-G. Meissner, E.~Oset and A.~Rusetsky,
\newblock Eur.Phys.J. {\bf A47}, 139 (2011), [arXiv:1107.3988].

\bibitem{Briceno:2012yi}
R.~A. Briceno and Z.~Davoudi,
\newblock [arXiv:1204.1110].

\bibitem{Rummukainen:1995vs}
K.~Rummukainen and S.~Gottlieb,
\newblock Nucl. Phys. B {\bf 450}, 397 (1995), [arXiv:hep-lat/9503028].

\bibitem{Kim:2005zzb}
C.~Kim, C.~T. Sachrajda and S.~R. Sharpe,
\newblock Nucl. Phys. B {\bf 727}, 218 (2005), [arXiv:hep-lat/0510022].

\bibitem{Bernard:2012bi}
V.~Bernard, D.~Hoja, U.~Meissner and A.~Rusetsky,
\newblock JHEP {\bf 1209}, 023 (2012), [arXiv:1205.4642].

\end{thebibliography}

\end{document}